\documentstyle[secnumtab]{fbssuppl}

\title{Hadron Probing of the Deuteron Structure
at Short Distances in Deuteron Breakup Reactions}

\author{A. P. Kobushkin\instnr{1}{\thanks{{\it E-mail address:}
akob@ap3.bitp.kiev.ua}}, A. P. Kostyuk\instnr{1} and
E. A. Eliseev\instnr{2}}
\instlist{Bogolyubov Institute for
Theoretical Physics, 252143, Kiev, Ukraine
\and Physics Department, Kiev National University,
Kiev, Ukraine}

\begin{document}

\maketitle
\begin{abstract}
We discuss deuteron $A(d,p)X$ breakup and cumulative pion production
$A(d,\pi)X$ in the framework of constituent quark model of the
deuteron.
We demonstrate that consideration of the Pauli principle at the
quark level, as well as multiple scattering, affect drastically cross
section and polarization observables of these reactions and provide
good description of the experimental data.
\end{abstract}

\begin{figure}[t]
\includegraphics{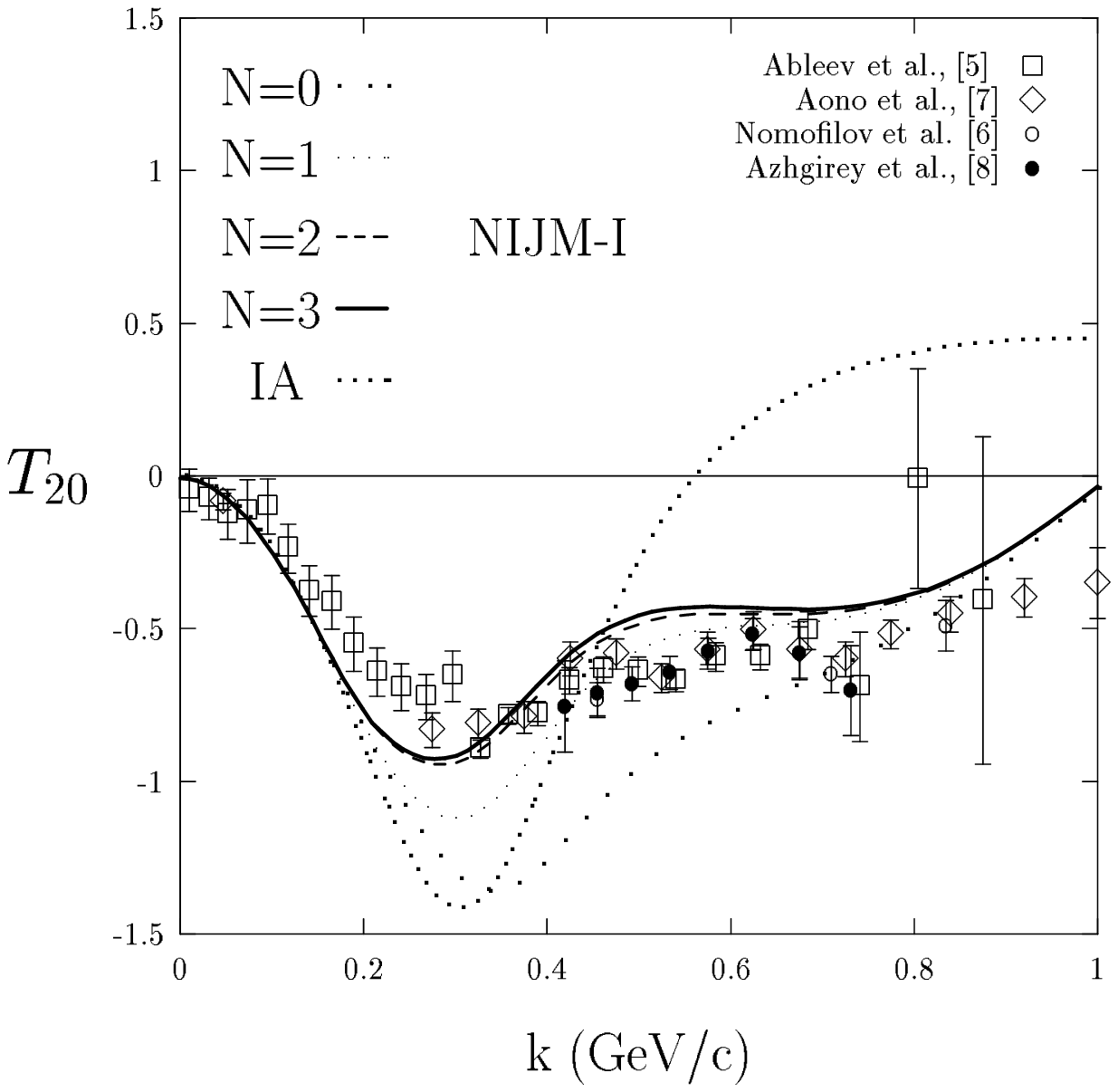}
\includegraphics{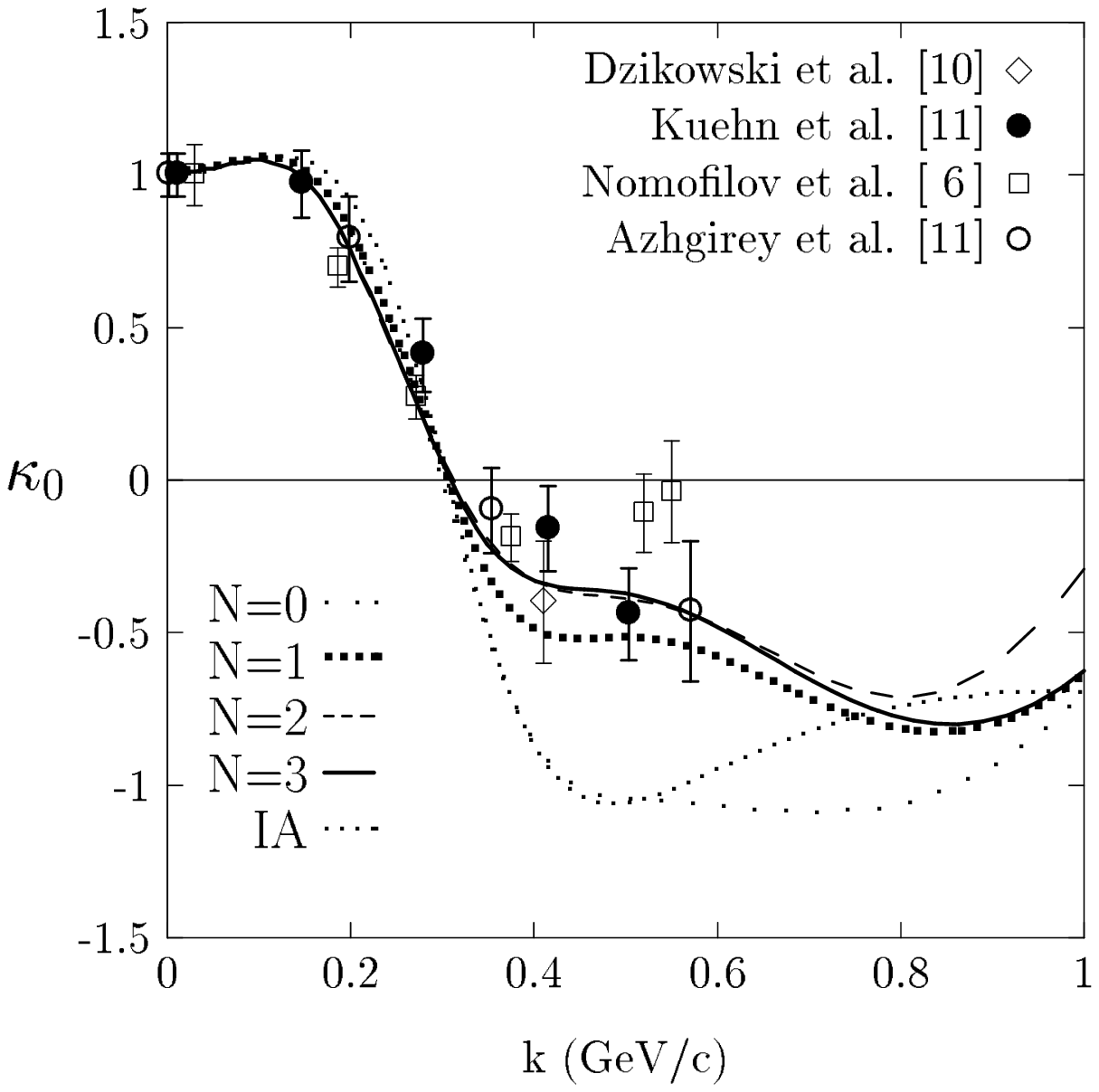}
\vspace{5.5cm}
\caption{The tensor analyzing power $T_{20}$ (left panel)
and the polarization transfer $\kappa_0$ (right panel) in the
$0^{\circ}$ inclusive $^{12}C(d,p)$ breakup as a function of the
relativistic internal momentum in the deuteron $k$. $N$ is
the maximum number of excitation quanta in the $N^{\ast}$ resonance in
the framework of oscillator quark model. The difference between $IA$
and $N=0$ demonstrates how important is multiple scattering.}
\end{figure}
\begin{figure}[b]
\includegraphics{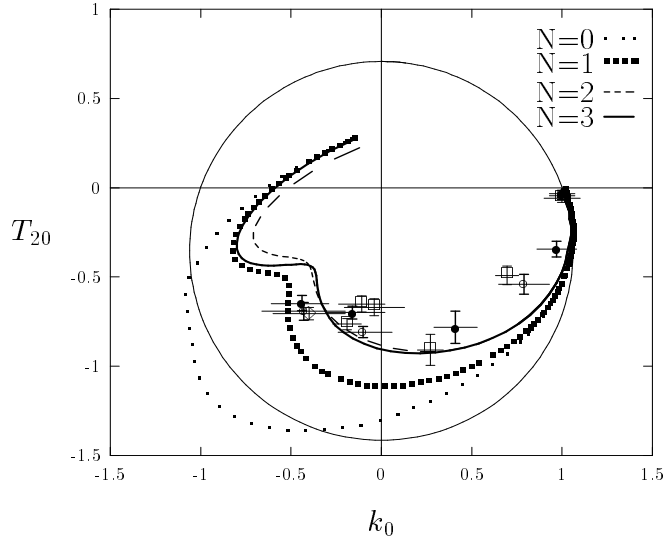}
\vspace{6.5cm}
\caption{The polarization observables of Fig. 1 on the KPS plot
[20].}
\end{figure}

During last 15 years the deuteron structure was studied in a wide
region, from that where description is given in terms of nucleons and
mesons to that where quarks and gluons should be explicitly used for the
deuteron description. Here we will analyze data on the deuteron breakup
$A(d,p)X$ \cite{Ableev83}--\cite{Kuehn} and cumulative pion production
$A(d,\pi)X$ \cite{Baldin,Afanasiev}
and show that both of them can be understood in the framework
of calculations which take into account the deuteron structure
generated by the Pauli principle at the constituent quark level.
In the present paper this was done using the simplest one-channel
approach of the Resonating Group Method (RGM) \cite{GBKK}
\begin{equation}
\psi^{d}(1,2,\ldots,6)= \hat{A}\left\{\varphi_{N}(1,2,3)
\varphi_{N}(4,5,6)\chi ({\bf r})\right\},
\label{rgm}
\end{equation}
where $\hat{A}$ is the quark antisymmetrizer and $\varphi_{N}$  are
wave functions of the  nucleon three quark ($3q$) clusters; $\chi
({\bf r})$ is the RGM distribution function and  $\bf r$ stands for
the relative coordinate between the centers of mass of the  $3q$
bags.

Due to the presence of the antisymmetrizer in Eq.(\ref{rgm}) the
deuteron wave function (DWF), being
de\-com\-posed into $3q \times 3q$ clusters, includes, apart from the
standard $pn$ component, nontrivial $\Delta \Delta$,
$NN^{\ast}$, $N^{\ast}N$ and $N^{\ast}N^{\ast\,\prime}$
components which correspond to all possible baryon resonance states
(see Ref.\cite{GBKK}). Most of the 1/2-isospin
isobars have negative parity and thus
generate effective $P$ waves of the DWF \cite{GBKK,KSG}. It
should be mentioned that any known quark approach to the DWF results in
admixture of $P$ components in the DWF \cite{GBKK,KSG,SPG} and study of
effects connected with them should give important information about the
deuteron structure at short distances.

Following Glozman-Kuchina paper (see Ref.\cite{GBKK}) we choose
$\chi(\bf r)$ as a conventional 
DWF, $\chi_{NN}(\bf r)$, modified
by the RGM renormalization condition of Ref.\cite{WT}.

Multiple scattering effects were incorporated in the framework of
the Bertocchi-Treleani model \cite{BT}. They appear to play an important
role especially at high internal momenta in the deuteron.

In Figs.~1 and 2 we compare results of our calculations with the
experimental data for the differential cross section, $T_{20}$
and $\kappa_0$ in the $A(d,p)X$ breakup.  The
results were obtained with Nijm-I potential \cite{deSwart}.

In spite of the fact that the $\Delta \Delta$ component has
the largest probability among the non-nucleonic components of the
deuteron, it does not contribute to $(d,p)$ breakup.
Nevertheless it may play significant role in cumulative pion production
$A(d,\pi)X$. In our calculations we use the following model:
the deuteron decays into virtual $\Delta \Delta$ pare, then one of the
$\Delta$-s is absorbed by the target and the other one yields the
cumulative pion and a nucleon. The results are
shown in Fig. 3.

We conclude that effects related to the Pauli principle at the level of
constituent quarks in the deuteron, as well as multiple scattering
effects, play crucial role in $A(d,p)X$ breakup, especially at high
deuteron internal momenta $k$. There is an evidence that the $\Delta
\Delta$ component in the deuteron provides a dominant contribution
into the cumulative pion production mechanism.

\begin{figure}[t]
\includegraphics{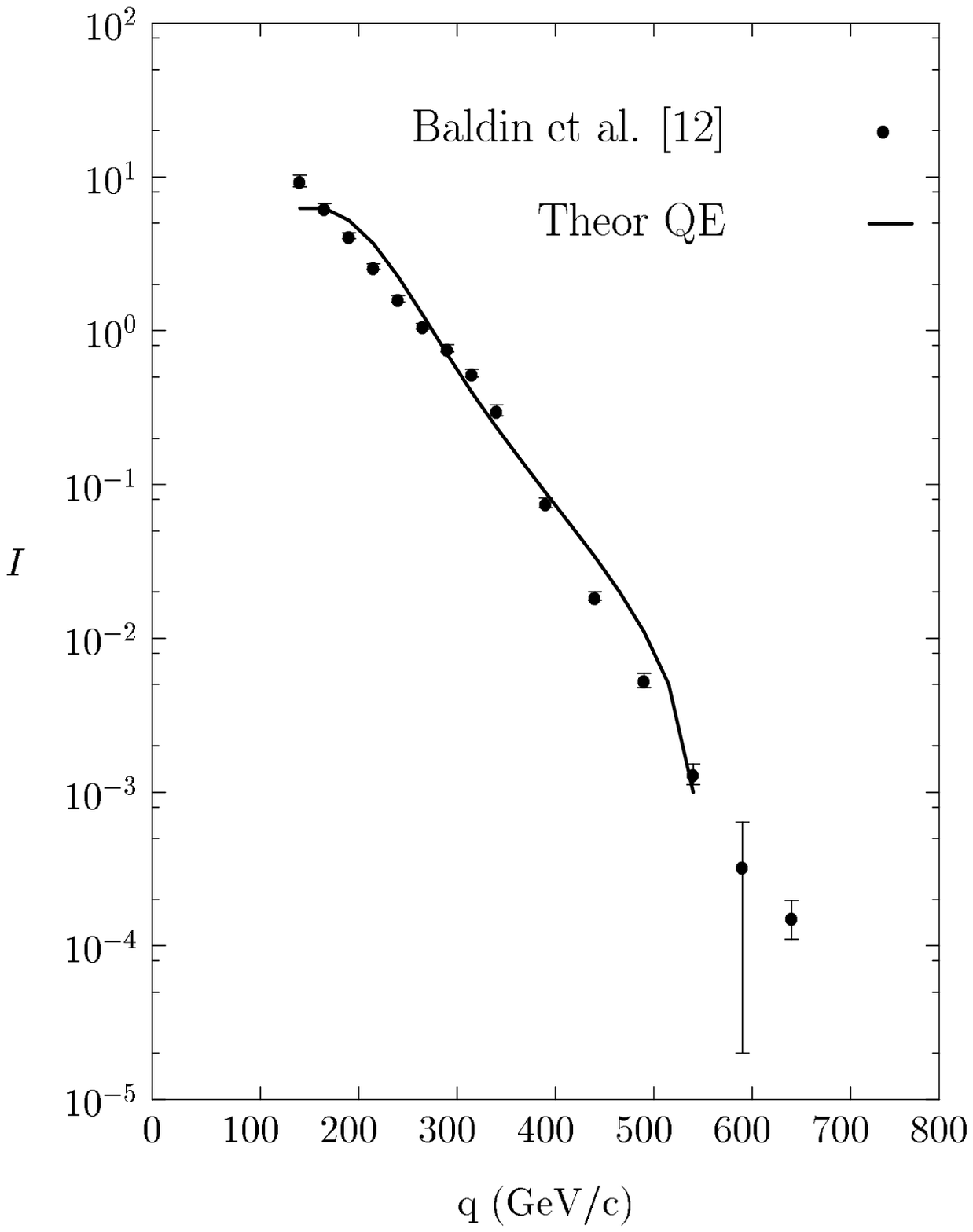}
\includegraphics{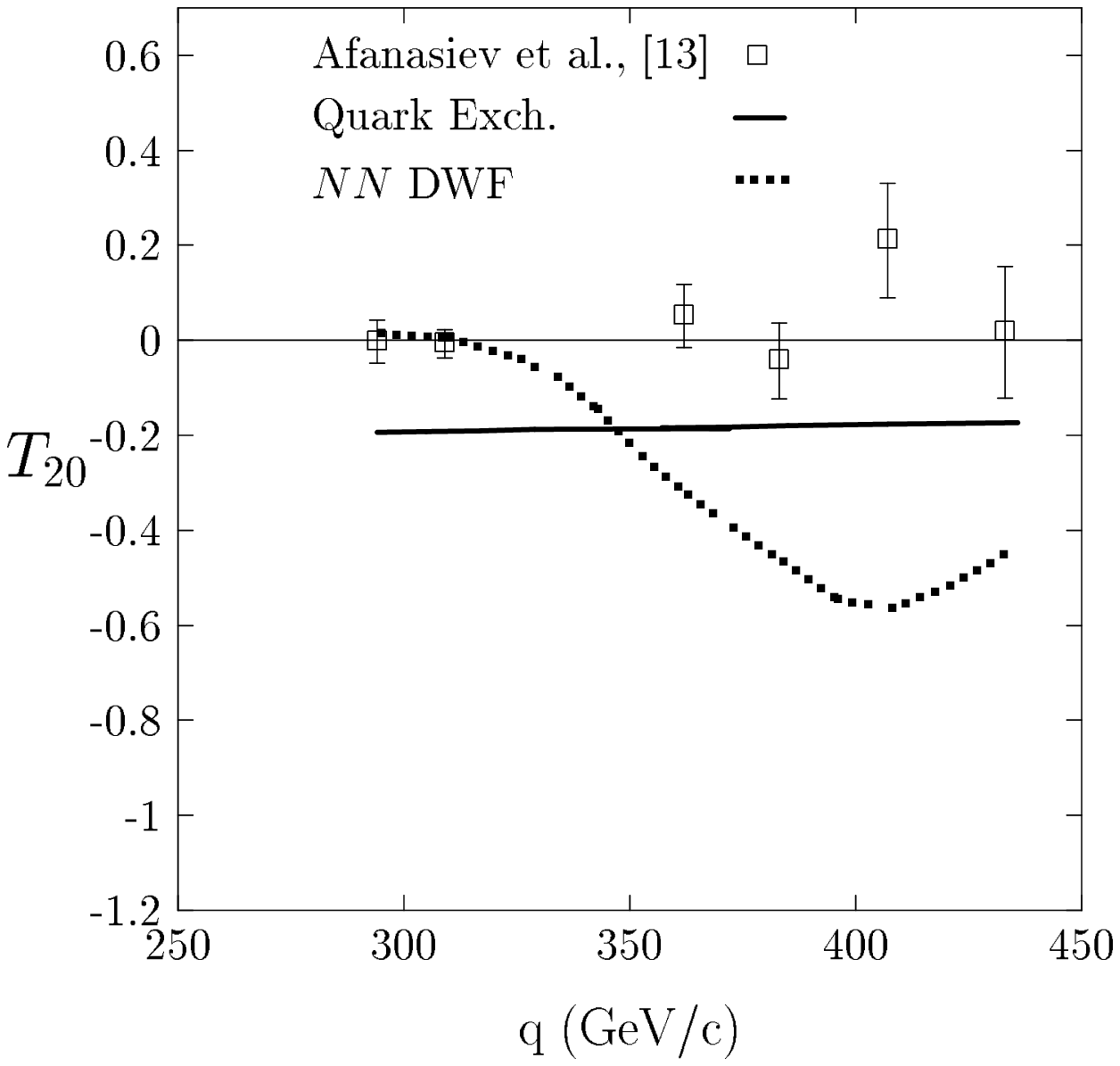}
\vspace{7.5cm}
\caption{The differential cross section
$I\equiv \frac{E}{2} \frac{d^3 \sigma}{d^3{p} }$
(left panel) and the
tensor analyzing power $T_{20}$ (right panel) of the $0^{\circ}$
cumulative pion production ($A(d,\pi)X$); $q$ is pion momentum in the
deuteron rest frame.}
\end{figure}

\begin{acknowledge}
One of us (A.P.Kobushkin) is grateful to Organizing Committee of
FB'98 for invitation and financial support during his stay in Autrans.
We thank C.Perdrisat for important comment.
\end{acknowledge}

\SaveFinalPage
\end{document}